\documentstyle[12pt]{article}
\renewcommand {\c}  {\'{c}}
\newcommand {\cc} {\v{c}}
\newcommand   {\s}  {\v{s}}
\begin{document}
\pagestyle{empty}
\vspace* {13mm}
\baselineskip=24pt
\begin{center}
%
  {\bf ON THE R-MATRIX FORMULATION OF DEFORMED ALGEBRAS AND
GENERALIZED JORDAN-WIGNER TRANSFORMATIONS}
  \\[20mm]

S.Meljanac$^1$, M.Milekovi\c$^{2,+}$ and A.Perica$^{1,++}$\\[7mm]
{\it $^1$ Rudjer Bo\s kovi\c \ Institute, Bijeni\cc ka c.54, 41001 Zagreb\\
Croatia\\[5mm]
 $^2$ Prirodoslovno-Matemati\cc ki Fakultet, Department of Theoretical
Physics, Bijeni\cc ka c.32, 41000 Zagreb\\Croatia\\[5mm]
$^+$ e-mail: marijan@phy.hr\\
$^{++} $ e-mail: perica@thphys.irb.hr}\\
\vskip 3cm
Classification number: $02.20 + b$;$03.65$
\end{center}
%
%
\newpage
\begin{center}
{\bf ABSTRACT}
\end{center}

The deformed algebra $\cal{A(R)}$, depending upon a Yang-Baxter R- matrix,
is considered. The conditions under which the algebra is
associative are discussed for a general number of oscillators. Four types of
solutions satisfying
these conditions are constructed and two of them can be
represented by generalized Jordan-Wigner transformations.Our analysis is
in some sense an extension of the boson realization of fermions from
single-mode
to  multimode oscillators.

%
%
%
\newpage
\setcounter{page}{1}
\pagestyle{plain}
\def\leer{\vspace{5mm}}
\setcounter{equation}{0}%
%
%
%
The introduction of q-deformations of the
Heisenberg-Weyl
algebras \cite{PW}
has led to the investigation of particles obeying the statistics
different from the ordinary Bose and Fermi statistics,
referred to as q-bosons and q-statistics,respectively. \\
Two particularly useful formulations of the q-deformed associative
Heisenberg-Weyl algebra are proposed through (i) constant solutions of
the Yang-Baxter-Hecke equations (R-matrices),which generalize
the notion of permutational symmetry \cite{FRT}, and (ii) generalized
Jordan-Wigner mappings from the undeformed Bose algebra \cite{FZ},
resembling  bosonization techniques widely used in nuclear physics
\cite{KM} and other fields of theoretical physics \cite{M}.
These two approaches may be considered as complementary in the sense that
there exist q-algebras which can be formulated in terms of R-matrices
but not as  regular mappings from the Bose algebra, and vice versa.

In a recent paper \cite{vdJ} the investigation was started with the algebra of
covariant
Pusz-Woronovicz oscillators and the corresponding $SU(n)_q$
R-matrix formulation of the deformed Bose algebra. Then, the same form of
deformed algebras
for a general  R-matrix was postulated. Under the requirements of associativity
and
hermiticity,  a set of conditions ( the Yang-Baxter and the Hecke equations )
was
obtained. These equations were solved for two oscillators  using an
ansatz that the R-matrices are of the eight-vertex type, and  three
types of solutions were found.

In this paper we construct four classes of deformed algebras for the
general case of n oscillators with the corresponding R-matrices.
The main characteristic of these algebras is whether or not they can
be obtained by continuous mapping from the Bose algebra.This can be
viewed as a generalization of the boson realization of fermions to multimode
oscillators of various types \cite{R}.

Let us start with the R-matrix deformed algebra \cite{vdJ} for n
oscillators

\begin{equation}
\begin{array}{c}
a_{i}a_{j} - p R_{ij,kl}  a_{l}a_{k}=0 \\[2mm]
a_{i}a^{+}_{j} - p' R_{ki,jl}  a^{+}_{k}a_{l}=\delta_{ij}
\end{array}
\end{equation}
where p and p' are real parameters and $ (i,j,k,l) \; {\epsilon} \;
(1,2,....n)$.
  The associativity and hermiticity requirements lead to the following
conditions:

(A) the Yang-Baxter equation\\
$$
\sum_{u,v,w}R_{ab,uv}R_{vw,cd}R_{ue,fw}=\sum_{u,v,w}R_{be,uv}R_{wu,fc}R_{av,wd}
$$

(B) the Hecke condition\\
$$
( p\hat{R} - 1)( p'\hat{R} +1 )=0
$$
where $\hat{R}$=PR and P is the permutation operator $P_{ij,kl} $= $\delta_{jk}
$$\delta_{il}$,
and the hermiticity requirement is $R_{ij,kl}=R^{*}_{lk,ji} $ or
$\hat{R}^{+}=\hat{R}$.

For general n, there are two types of solutions given by

${\bf (i)}$ $\qquad$ $p\hat{R}=1 $ $\qquad$ i.e. $\qquad$ $ pR=P $\\
with the corresponding algebra

\begin{equation}
a_{i}a^{+}_{j}=Q\delta_{ij}
\end{equation}
where
\begin{equation}
Q=1+\frac {p'}{p}\sum^{n}_{i=1}a^{+}_{i}a_{i}
\end{equation}

${\bf (ii)}$ $\qquad$ $p'\hat{R}=-1$  $\qquad$ i.e.  $\qquad$ $p'R=-P$\\
with the corresponding algebra

\begin{equation}
\begin{array}{c}
a_{i}a^{+}_{j}=Q\delta_{ij}\\[2mm]
a_{i}a_{j}=0
\end{array}
\end{equation}
where

\begin{equation}
Q=1-\sum^{n}_{i=1}a^{+}_{i}a_{i}
\end{equation}

The main characteristic of solutions (i) and (ii) is that there exists no
mapping to the Bose algebra .
In  case (i)  there are no commutation relations between the annihilation
$a_{i}$,
 $a_{j}$ ( creation $a^{+}_{i}$, $a^{+}_{j}$ ) operators.  The $a_{i}$, $a_{j}$
operators behave
as some kind of generalized quons and, in the limit $\frac {p'}{p}$=$0$, they
are the same as quons with $q=0$ \cite{Gr}. In case (ii), the oscillators
behave as
objects with the hard-core condition $a^{2}_{i}$=$0$. This is in some sense
similar to the fermionic algebra.The algebra (4) is described by Klein and
Marshalek in refs. \cite{KM} \cite{KM1}.\\
In addition to solutions (i) and (ii) there are other two types of
solutions corresponding  to oscillators that can be obtained by mappings
from Bose oscillators in the form of generalized Jordan-Wigner
transformations.

${\bf (iii)}$ One mapping leading to the q-deformed algebra (1) is of the form

\begin{equation}
a_{i}=[\prod^{n}_{j=1}(\kappa_{ij}q)^{\theta_{ij}N_{j}}]
b_{i}\sqrt{\frac {[N_{i}]_{\chi_{i}}}{N_{i}}}
\end{equation}
where
$$
[N_{i}]_{\chi_{i}}=\frac {\chi^{N_{i}}_{i}-1}{\chi_{i}-1}\\[2mm]
$$
\begin{equation}
\chi_{i}=\frac{q^2-1}{2}+\epsilon_{i}\frac{q^2+1}{2}\\[2mm]
\end{equation}
$$
\kappa_{ij}=\kappa_{ji} \in \{-1,1\}\\[2mm]
$$
$$
\epsilon_{i} \in \{-1,1\} ,  \quad q \in {\bf R}^{+}
$$
and $N_i$ is the number operator for the i-th oscillator mode.\\
The corresponding deformed algebra is explicitly given by

\begin{equation}
\begin{array}{c}
(\epsilon_{i} -1) a^{2}_{i}=0\\[2mm]
a_{i}a_{j}=\kappa_{ij} q^{sgn(j-i)} a_{j}a_{i}\\[2mm]
a_{i}a^{+}_{j}=\kappa_{ij} q a^{+}_{j}a_{i}, \qquad  i \neq j \\[2mm]
\end{array}
\end{equation}
$$
a_{i}a^{+}_{i}=q^{2\sum_{j}\theta_{ij}N_{j}}+\chi_{i}a^{+}_{i}a_{i}=\\[2mm]
$$
$$
=1+(q^{2}-1)\sum_{j}\theta_{ij}a^{+}_{j}a_{j}+\chi_{i}a^{+}_{i}a_{i}
$$
where $\theta_{ij}$ is a step-function.\\
The corresponding R-matrix is of the form
\begin{equation}
p'R=\sum_{i}\chi_{i} e_{ii} \otimes e_{ii}+q\sum_{i\neq j}\kappa_{ij} e_{ii}
\otimes e_{jj}+(q^{2}-1)\sum_{i<j}e_{ij}\otimes e_{ji}=q^{2}  pR
\end{equation}
where $(e_{ij})_{kl}=\delta_{ik}\delta_{jl}$.

In solution (iii) there are ( in addition to ordinary bosons, fermions and
Green's oscillators for para-Bose and para-Fermi statistics \cite{GreenOK} )
covariant Pusz-Woronowicz oscillators of both the bosonic and the fermionic
type. This is a generalization of the results in ref. \cite{R} to  multimode
oscillators.The algebra in eq.(8) is automatically
associative and norms of the states in the Fock space are positive definite.

${\bf (iv)}$ There is also another mapping leading to the q-deformed algebra
(1):

\begin{equation}
a_{i}=e^{i \sum_{j}(\lambda_{ij} - \lambda_{ji})N_{j}}b_{i}
\sqrt{\frac {[N_{i}]_{\xi_{i}}}{N_{i}}}
\end{equation}
with $\lambda_{ij}$  real numbers and  $\xi_{i}=\pm 1$.
The oscillators $ a_{i} $
are of the anyonic type in the single-valued picture in $(2+1)$ dimensions,
where $\lambda_{ij}$ becomes the angle function
\cite{FZ}\cite{BDMLS}\cite {LS}.
They satisfy the following algebra:

\begin{equation}
\begin{array}{c}
(\xi_{i} -1) a^{2}_{i}=0\\[2mm]
a_{i}a_{j}= e^{i(\lambda_{ji}-\lambda_{ij})} a_{j}a_{i} \qquad  i \neq j
\\[2mm]
a_{i}a^{+}_{j}=e^{i(\lambda_{ij}-\lambda_{ji})}a^{+}_{j}a_{i}, \qquad  i \neq j
\\[2mm]
a_{i}a^{+}_{i}=1+ \xi_{i} a^{+}_{i}a_{i}
\end{array}
\end{equation}

$\xi_{i}=1$ ($\xi_{i}=-1$) corresponds to the Bose - ( Fermi- ) type
commutation relation.The above algebra can be written in the R-matrix
formulation,eq.(1), only for $\xi_{i}= \pm 1.$\\
Then,the corresponding R-matrix is
\begin{equation}
p'R=\sum_{i}\xi_{i} e_{ii} \otimes e_{ii}+\sum_{i\neq j} e^{i(\lambda_{ij} -
\lambda_{ji})} e_{ii}
\otimes e_{jj}=pR
\end{equation}
or
\begin{equation}
(pR)_{ij,kl}=e^{i(\lambda_{ij}- \lambda_{ji})}\delta_{ik}
\delta_{jl}+(\xi_{i}-1)
\delta_{ij}\delta_{ik}\delta_{jl}
\end{equation}
This R-matrix can be written in the continuum limit leading to the
q-deformed field theory in $(2+1)$ dimensions \cite{BDMLS}.
Pure anyons are characterized by $\lambda_{ij}- \lambda_{ji} = \lambda
\pi sgn(i-j)$,with $\lambda = const.$
We point out that multivalued anyons in $2+1$ dimensions \cite{LMW}
can also be written in the form of multivalued mapping to bosons \cite{BDM}
and the corresponding algebra is given by \\

\begin{equation}
\begin{array}{c}
a_{i}a_{j}=e^{i\lambda \Delta}a_{j}a_{i}\\[2mm]
a_{i}a^{+}_{j}=\delta_{ij} +e^{i\lambda \Delta}a^{+}_{j}a_{i}\\[2mm]
\Delta =\{ \pi (1+2z); z \in Z\}
\end{array}
\end{equation}
and the R-matrix is

\begin{equation}
(pR)_{ij,kl}=(p'R)_{ij,kl}=e^{i\lambda \Delta}\delta_{ik}\delta_{jl}
\end{equation}
where the R-matrix elements are multivalued numbers. ( The
multiplication "unit" is of the form $E=e^{2ik\pi\lambda}, k\epsilon {\bf Z}$
).

Our solution (iv) of the anyonic type for $n=2$ was overlooked in \cite{vdJ}.
Adding this solution to the solutions in  \cite{vdJ} makes the list of
solutions for
$n=2$ complete.This can be checked by inspection of the complete list of
solutions to the Yang-Baxter equations for $n=2$ \cite{Hiet}.
Finally, there may exist other solutions with "peculiar" statistics for which
no mapping
to the Bose algebra exists ( for $n=2$ see \cite{vdJ} ) and the number
operators do
not exist in the usual sense.
Owing to their "peculiarity",it is
interesting to discuss the Fock-space representation of "peculiar" algebras
 and to construct them for a general number of oscillators.The results of
this investigation are reported elsewhere \cite{MMP}.
%
%
\newpage
{\bf Acknowledgments}

This work was supported by the joint Croatian-American contract NSF JF 999 and
the Scientific Fund of the Republic of Croatia.
%
%
%

%
%
%
%
%
\end{document}